\documentclass[pdflatex,sn-standardnature]{sn-jnl}

\newcommand{\araa}{Annu. Rev. Astron. Astrophys.}   
\newcommand{\aj}{Astron. J.}   
\newcommand{\apj}{Astrophys. J.}   
\newcommand{\apjl}{Astrophys. J. Lett.}   
\newcommand{\apjs}{Astrophys. J. Suppl. Ser.}   
\newcommand{\aap}{Astron. Astrophys.}   
\newcommand{\mnras}{Mon. Not. R. Astron. Soc.}   
\newcommand{\nat}{Nature} 
\newcommand{\nastro}{Nat. Astron.} 
\newcommand{\pasa}{Publ. Astron. Soc. Aust.}   
\newcommand{\pasp}{Publ. Astron. Soc. Pac.}   

\jyear{2021}%

\theoremstyle{thmstyleone}%
\theoremstyle{thmstyletwo}%

\theoremstyle{thmstylethree}%

\newcommand\rev[1]{{ #1}} 
\newcommand\revv[1]{{ #1}} 
\newcommand\revvv[1]{{  #1}} 

\begin{document}

\title[Nested Dust Shells around the Wolf-Rayet Binary WR~140 observed with JWST]{Nested Dust Shells around the Wolf-Rayet Binary WR~140 observed with JWST}


\author*[1]{Ryan M. Lau}\email{ryan.lau@noirlab.edu}
\author[2]{Matthew J. Hankins}
\author[3]{Yinuo Han}
\author[4]{Ioannis Argyriou}
\author[5,6]{Michael F. Corcoran}
\author[7]{Jan J. Eldridge}
\author[8]{Izumi Endo}
\author[9]{Ori D. Fox}
\author[10]{Macarena Garcia Marin}
\author[9,11]{Theodore R. Gull}
\author[12]{Olivia C. Jones}
\author[5,13]{Kenji Hamaguchi}
\author[14,15]{Astrid Lamberts}
\author[9]{David R. Law}
\author[16]{Thomas Madura}
\author[17]{Sergey V. Marchenko}
\author[18]{Hideo Matsuhara}
\author[19]{Anthony F.J. Moffat}
\author[20]{Mark R. Morris}
\author[21]{Patrick W. Morris}
\author[8,22]{Takashi Onaka}
\author[23]{Michael E. Ressler} 
\author[24]{Noel D. Richardson}
\author[25]{Christopher M. P. Russell}
\author[26,27]{Joel Sanchez-Bermudez}
\author[28]{Nathan Smith}
\author[29]{Anthony Soulain}
\author[30]{Ian R. Stevens}
\author[31]{Peter Tuthill}
\author[32]{Gerd Weigelt}
\author[33]{Peredur M. Williams\textsuperscript{33}}
\author[34,18]{Ryodai Yamaguchi}

\affil*[1]{NSF’s NOIRLab, 950 N. Cherry Avenue, Tucson, 85719, AZ,
USA}

\affil[2]{Arkansas Tech University,  215 West O Street, Russellville, AR 72801, USA}

\affil[3]{Institute of Astronomy, University of Cambridge, Madingley Road, Cambridge CB3 0HA, UK}

\affil[4]{Instituut voor Sterrenkunde, KU Leuven, Celestijnenlaan 200D, bus-2410, 3000 Leuven, Belgium}

\affil[5]{CRESST II and X-ray Astrophysics Laboratory NASA/GSFC, Greenbelt, MD 20771, USA}

\affil[6]{Institute for Astrophysics and Computational Sciences, The Catholic University of America, 620 Michigan Ave., N.E. Washington, DC 20064, USA}

\affil[7]{Department of Physics, University of Auckland, Private Bag 92019, Auckland, New Zealand}

\affil[8]{Department of Astronomy, School of Science, University of Tokyo, 7-3-1 Hongo, Bunkyo-ku, Tokyo 113-0033, Japan}

\affil[9]{Space Telescope Science Institute, 3700 San Martin Dr., Baltimore, MD 21218, USA}

\affil[10]{European Space Agency, Space Telescope and Science Institute, 3700 San Martin Drive, Baltimore, MD, USA}

\affil[11]{Code 667, NASA/GSFC, Greenbelt, MD 20771}

\affil[12]{UK Astronomy Technology Centre, Royal Observatory, Blackford Hill, Edinburgh, EH9 3HJ, UK}

\affil[13]{Department of Physics, University of Maryland, Baltimore County, 1000 Hilltop Circle, Baltimore, MD 21250, USA}

\affil[14]{Université Côte d’Azur, Observatoire de la Côte d’Azur, CNRS, Laboratoire Lagrange, Bd de l’Observatoire,
CS 34229, 06304 Nice cedex 4, France}

\affil[15]{Université Côte d’Azur, Observatoire de la Côte d’Azur, CNRS, Laboratoire Artémis, Bd de l’Observatoire, CS 34229, 06304 Nice cedex 4, France}

\affil[16]{Department of Physics and Astronomy, San Jose State University, One Washington Square, San Jose, CA 95192, USA}

\affil[17]{Science Systems and Applications, Inc., Lanham, MD, USA and NASA/GSFC, Greenbelt, MD, USA}

\affil[18]{Institute of Space and Astronautical Science, Japan Aerospace Exploration Agency, 3-1-1 Yoshinodai, Chuo-ku, Sagamihara, Kanagawa 252-5210 Japan}

\affil[19]{Dépt. de physique, Univ. de Montréal, CP 6128, Succ. C-V, Montréal, QC H3C 3J7 and Centre de Recherche en Astrophysique du Québec, Canada}

\affil[20]{Department of Physics and Astronomy, University of California, Los Angeles, CA 90095-1547, USA}

\affil[21]{Caltech/IPAC, Mailcode 100-22, Pasadena, CA 91125, USA}

\affil[22]{Department of Physics, Faculty of Science and Engineering, Meisei University, 2-1-1 Hodokubo, Hino, Tokyo 191-8506, Japan}

\affil[23]{Jet Propulsion Laboratory, California Institute of Technology, 4800 Oak Grove Drive, Pasadena, CA 91109, USA}

\affil[24]{Department of Physics and Astronomy, Embry-Riddle Aeronautical University, 3700 Willow Creek Rd, Prescott, AZ, 86301, USA}

\affil[25]{Department of Physics and Astronomy, Bartol Research Institute, University of Delaware, Newark, DE 19716 USA}

\affil[26]{Instituto de Astronom\'ia, Universidad Nacional Aut\'onoma de M\'exico, Apdo. Postal 70264, Ciudad de M\'exico, 04510, M\'exico}

\affil[27]{Max-Planck-Institut f\"ur Astronomie, K\"onigstuhl 17, D-69117 Heidelberg, Germany}

\affil[28]{Department of Astronomy, University of Arizona, 933 N. Cherry St, Tucson, AZ 85721, USA}

\affil[29]{Univ. Grenoble Alpes, CNRS, IPAG, 38100 Grenoble, France}

\affil[30]{School of Physics and Astronomy, University of Birmingham, Birmingham B15 2TT, UK}

\affil[31]{Sydney Institute for Astronomy, School of Physics, University of Sydney, NSW 2006, Australia}

\affil[32]{Max Planck Institute for Radio Astronomy, Auf dem H\"ugel 69, 53121 Bonn, Germany}

\affil[33]{Institute for Astronomy, University of Edinburgh, Royal Observatory,  Blackford Hill, Edinburgh, EH9 3HJ, UK}

\affil[34]{Graduate School of Science and Technology, Tokyo Institute of Technology, Meguro-ku, Tokyo 152-8551, Japan}


\abstract{\revv{Massive colliding-wind binaries that host a Wolf-Rayet (WR) star present a potentially important source of dust and chemical enrichment in the interstellar medium (ISM). However, the chemical composition and survival of dust formed from such systems is not well understood. The carbon-rich WR (WC) binary WR~140 presents an ideal astrophysical laboratory for investigating these questions given its well-defined orbital period and predictable dust-formation episodes every 7.93 years around periastron passage.}  
We present observations from our Early Release Science program (ERS1349) with the James Webb Space Telescope (JWST) Mid-Infrared Instrument (MIRI) Medium-Resolution Spectrometer (MRS) and Imager that reveal the spectral and spatial \revv{signatures} of nested circumstellar dust shells around WR~140.
MIRI MRS spectroscopy of the second dust shell and Imager detections of over 17 shells \revv{formed throughout the past $\gtrsim130$ years} confirm the survival of carbonaceous \rev{dust grains} from WR~140 that are likely carriers of ``unidentified infrared" (UIR)-band features at 6.4 and 7.7 $\mu$m. The observations \revv{indicate} that dust-forming WC binaries \revv{can enrich the ISM with} organic compounds and carbonaceous \rev{dust}. }






\maketitle

\section{Introduction}
\label{sec:intro}

Some of the first carbonaceous dust grains and organic material in the Universe may have been produced in the compressed, chemically-enriched winds of massive, evolved colliding-wind binaries hosting a carbon-rich Wolf-Rayet (WC) star \citep{Marchenko2017,Lau2020a,Endo2022}.  In the hostile environment around these massive and evolved stars \revv{\citep{Crowther2007}}, dust is produced from the compressed gas in the wind collision region where the stronger stellar wind from the Wolf-Rayet (WR) star interacts with the stellar wind of an OB-star companion \citep{Usov1991,Williams2009,Eatson2022}. Galactic colliding-wind WC binaries with resolvable circumstellar dust nebulae therefore provide important laboratories to study this dust-formation process, where observations over the past few decades have demonstrated how dust formation is regulated by the orbit of the binary system \citep{Usov1991,Tuthill1999,Williams2009,Lau2020a}.
However, the chemical signatures of dust grains formed in colliding-wind WC binaries \citep{Cherchneff2000} and whether or not the grains survive as they propagate into the interstellar medium (ISM) in their harsh radiative stellar environment \rev{are} currently not well understood.  

Arguably the best example of ``episodic'' dust formation in colliding winds is the Wolf-Rayet binary WR~140 (also known as HD 193793), a carbon-rich evolved massive star of subtype WC7 in a highly-eccentric (e = 0.89), 2895-day (7.93-year) mutual orbit around an early-type O5.5fc star \citep{Monnier2011,Fahed2011, Thomas2021}. Unlike \rev{continuously} dust-forming WC binaries like WR~104 \citep{Tuthill1999} that are known for their ``pinwheel'' nebulae\revv{,} WR~140 periodically produces dust over a few months duration and exhibits a nebular morphology resembling segmented shells \citep{Williams2009}. This episodic dust production by WR~140 occurs when the dense stellar wind of the WC7 star is sufficiently compressed around periastron passage by the wind of the O star \citep{Williams2009,Monnier2011,Eatson2022}. The dust condenses in the shock-compressed WC wind and is carried into the circumstellar medium, likely seeding the ISM with new, carbonaceous dust. 
\rev{Given the relatively close distance of 1.64 kpc to WR~140 \citep{Rate2020,Thomas2021},}
previous IR imaging by ground-based telescopes \rev{has} resolved \rev{thermal emission from} (at most) two discrete, nested dust shells \rev{out to $\sim5000$ AU from the central binary and has} provided a comparison of dust properties from one periastron passage to the next \citep{Williams2009}. 
However, mid-IR spectroscopy of spatially resolved dust emission and sensitive imaging observations detecting more distant dust shells are critical for investigating the chemical composition and survival of WC dust. Such observations have been practically impossible due to sensitivity limitations of ground-based facilities and low angular resolution of previous space-based observatories. 

As part of the James Webb Space Telescope (JWST) Director’s Discretionary Early Release Science (DD-ERS) program (PID - ERS1349), we observed WR~140 with JWST to investigate the chemical composition and survival of dust formed by colliding winds of WC binaries.
WR~140 also presented an ideal target for the DD-ERS program to demonstrate the capabilities of JWST/MIRI for resolving and detecting faint extended emission around a bright central point source. In this paper, we discuss the JWST/MIRI observations that reveal the spectral and spatial \revv{signatures} of dust around WR~140: 
over 17 nested dust shells from the past $\gtrsim130$ yr of episodic dust production and the mid-IR spectroscopic signatures from the direct emission of WC dust. 

\begin{figure*}[t!]
    \includegraphics[width=.98\linewidth]{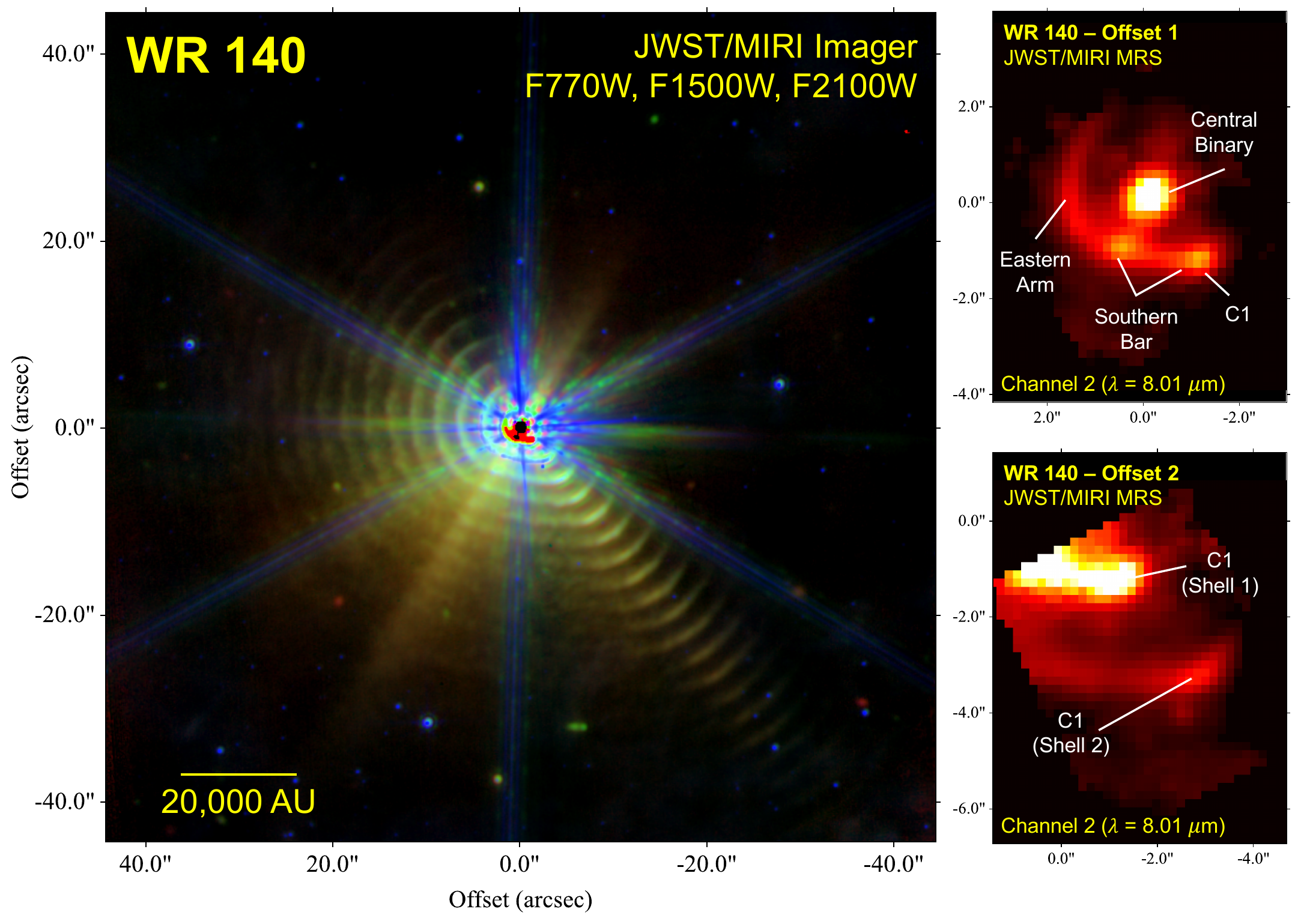}    \caption{ \textbf{JWST/MIRI Observations of WR~140.} (Left) False color JWST/MIRI Imager observations of WR~140 taken with the F770W, F1500W, and F2100W filters that correspond to blue, green, and red, respectively. \rev{Eight symmetric diffraction spikes are seen around the saturated core of WR~140 (at origin) and exhibit bluer colors than the dust emission.}
    (Right) MIRI MRS spectral data cube images of WR~140 from Channel 2 at 8.01 $\mu$m taken at two different positions that cover the central binary and shell 1 (Offset 1) and shells 1 and 2 (Offset 2) shown in a square-root stretch. \rev{The central binary and bright dust emission features are labelled in the MRS data cube images.}}
    \label{fig:WR140Image}
\end{figure*}

\section{Results and Discussion}

WR~140 was observed by JWST with the Mid-Infrared Instrument (MIRI, \citep{Bouchet2015,Wright2015,Wells2015}) Medium-Resolution Spectrometer (MRS) and Imager on 2022 July 8 UT and 2022 July 27 UT, respectively. The timing of the observations captures the binary system at an orbital phase of 0.7, about $\sim5.5$ yr after its last \rev{dust-formation episode} and periastron passage on 2016 Dec \revv{\citep{Thomas2021}}.
These MIRI Imager (Fig.~\ref{fig:WR140Image}, \textit{Left}) and MRS (Fig.~\ref{fig:WR140Image}, \textit{Right}) observations provide the most sensitive mid-infrared ($5-28$ $\mu$m) imaging and spectroscopy of WR~140 and its circumstellar environment to date \rev{by over two orders of magnitude}. 
The MIRI Imager observations of WR~140 at 7.7, 15, and 21 $\mu$m shown in Fig.~\ref{fig:WR140Image} (\textit{Left}) reveal thermal emission from nested and remarkably evenly spaced circumstellar dust shells \rev{exceeding $\sim45''$ ($\sim70000$ AU) from the central binary.} 
\revvv{The extent of the these distant circumstellar shells detected around WR~140 exceeds that of all other known dust-forming WC systems by factors of 4 or greater \citep{Lau2020a,Lau2020b,Han2020}.}

The central binary and the two most recent dust shells in the Imager observations are affected by detector saturation and/or partially overlap with the point-spread function (PSF) of the bright core, which is dominated by free-free emission from ionized stellar winds \citep{Williams2009}. However, due to the dispersion of incoming light, integral field unit (IFU) imaging of these inner regions with the MRS were not affected by saturation (Fig.~\ref{fig:WR140Image} \textit{Right}).

\begin{figure*}[t!]
    \includegraphics[width=.98\linewidth]{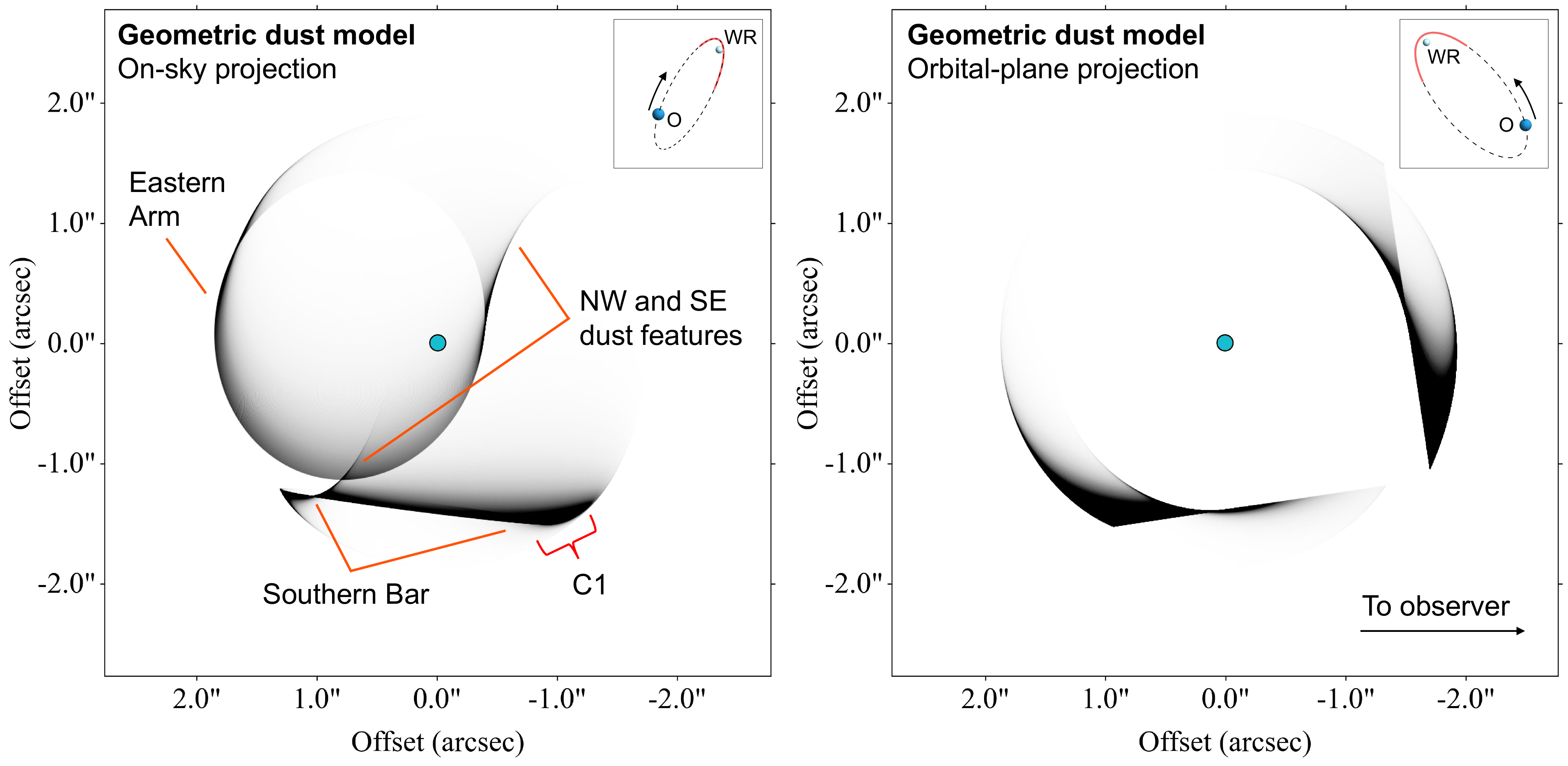}    \caption{ \rev{\textbf{Views of the single-shell geometric dust model of WR~140 with on-sky and orbital-plane projections at the orbital configuration during MIRI observations.} (Left) Dust column density in the on-sky projection of the geometric model, labelled with the observed dust features. (Right) Dust column density in the orbital-plane projection of the model. In both (Left) and (Right), the cyan dot corresponds to the location of the central binary. The projected orbital configuration of WR~140 in the reference frame of the WR star and the direction of motion of the O-star is shown in the upper-right insets from each projection. The semi-major axis of the orbit is 8.9 mas (14.6 AU) \citep{Thomas2021}. The red regions in \revv{the} orbital schematic \revv{around periastron passage correspond to true anomalies spanning $-135^\circ$ and $+135^\circ$} where dust forms from the colliding winds \revv{(Sec.~\ref{sec:model}). Dust formation does not occur at a constant rate throughout periastron passage: more dust is formed from colliding winds around the beginning and end of the orbital regions outlined in red \citep{Han2022}.}}}
    \label{fig:WR140Orbit}
\end{figure*}

\revv{One of the} brightest repeating circumstellar dust \revv{features} \rev{in the MIRI \revv{imaging} observations} is associated with the south-western `C1' dust emission peak\revv{, which is labeled in the MIRI MRS spectral data cube image at 8.01 $\mu$m shown in (Fig.~\ref{fig:WR140Image}, \textit{Right})}. \revv{The C1 feature} was initially identified from ground-based mid-IR imaging \revv{of WR~140} by \citep{Williams2009}.
\rev{The horizontal ``Southern Bar,'' which extends east-ward from C1, and the curved ``Eastern Arm'' are also previously identified features that show repeating structures in the MIRI \revv{imaging} observations. However, the Southern Bar and Eastern Arm are not detected as far out from the central binary as C1.}
Eight prominent \rev{and symmetrical} diffraction spikes due to the secondary struts and the hexagonal shape of JWST's primary mirror are \rev{present around the saturated core of WR~140} in the \rev{MIRI} image \rev{and exhibit bluer colors than the dust emission (Fig.~\ref{fig:WR140Image} \textit{Left})}.
Interestingly, there are two additional asymmetric and `redder' linear features that extend from the core of WR~140 to the north-west and south-east. These \rev{linear NW and SE} features are not consistent with known instrumental artefacts from bright point sources, which \rev{indicates} that their origin is astrophysical and likely \rev{attributable} to emission from circumstellar dust.

\begin{figure*}[t!]
    \includegraphics[width=.98\linewidth]{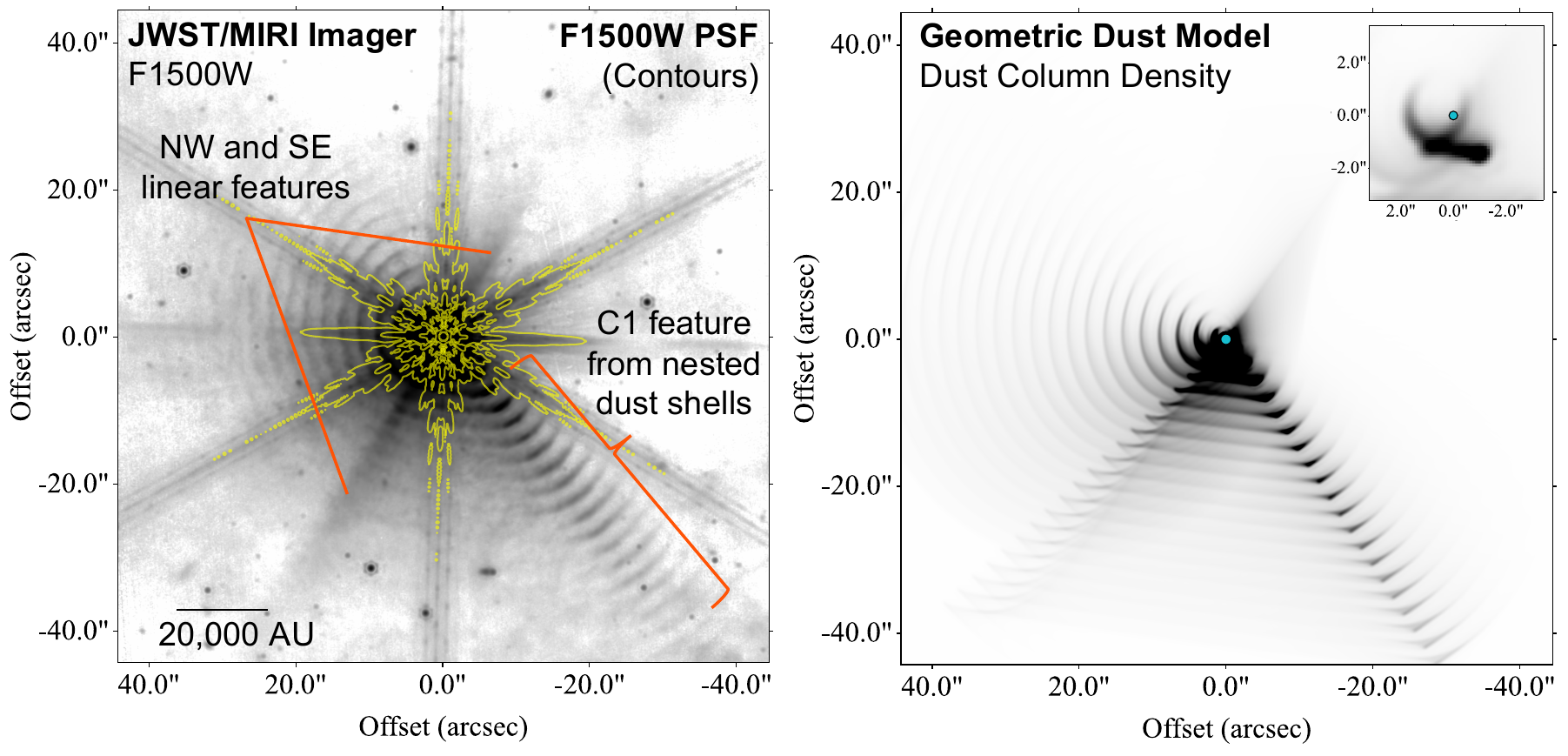}    \caption{ \textbf{Comparison of MIRI F1500W image of WR~140 with \rev{the 20-shell} geometric dust model.} (Left) MIRI Imager observations of WR~140 taken with the F1500W filter. \rev{\revv{Yellow} contours show the diffraction spikes from the MIRI F1500W PSF generated by WebbPSF \citep{Perrin2014} at levels corresponding to [$10^{-5}$, $10^{-4}$, $10^{-3}$, $10^{-2}$] times the peak intensity of the PSF.} (Right) \rev{Dust} column density model convolved with the MIRI F1500W PSF. The inset in the upper right shows the inner $\sim3''$ of the dust model with the central core. \rev{The cyan dot indicates the position of the central binary.}}
    \label{fig:WR140Mod}
\end{figure*}

The observed morphology of the circumstellar dust emission \rev{around WR~140} \rev{can be interpreted by} a geometric dust shell model (Sec.~\ref{sec:model}). \rev{Models were} generated at an orbital phase consistent with the JWST observations \rev{to simulate} the dust column density of \rev{a single dust shell (Fig.~\ref{fig:WR140Orbit})} \rev{as well as} 20 nested dust shells (Fig.~\ref{fig:WR140Mod}, \textit{Right}). \rev{These} geometric dust shell \rev{models} utilize the well-constrained measurements of WR~140's orbital parameters \citep{Monnier2011,Thomas2021} and \rev{have} been previously used to interpret the expansion of newly formed dust around periastron passage \citep{Han2022}.
\rev{The single-shell model (Fig.~\ref{fig:WR140Orbit}) is shown in both on-sky and orbital-plane projections and demonstrates how column density enhancements along the line-of-sight reproduce the bright dust emission features seen in the MRS observations (Fig.~\ref{fig:WR140Image} \textit{Right}). Specifically, the C1 feature is composed of dust along the projected southwestern edge of the shell. Similarly, the Eastern Arm and Southern Bar arise due to enhanced column densities along the projected eastern and southern edges of the shell, respectively.}
\revv{Note that when WR~140 forms dust, it does not do so at a constant rate \citep{Han2022}. Enhanced dust formation occurs from colliding winds around the beginning and end of periastron passage, which explains the relative absence of dust to the northwest of WR~140 (Fig.~\ref{fig:WR140Image}~\&~\ref{fig:WR140Orbit}, See Sec.~\ref{sec:model}).}

\begin{figure}[t!]
    \centerline{\includegraphics[width=0.7\linewidth]{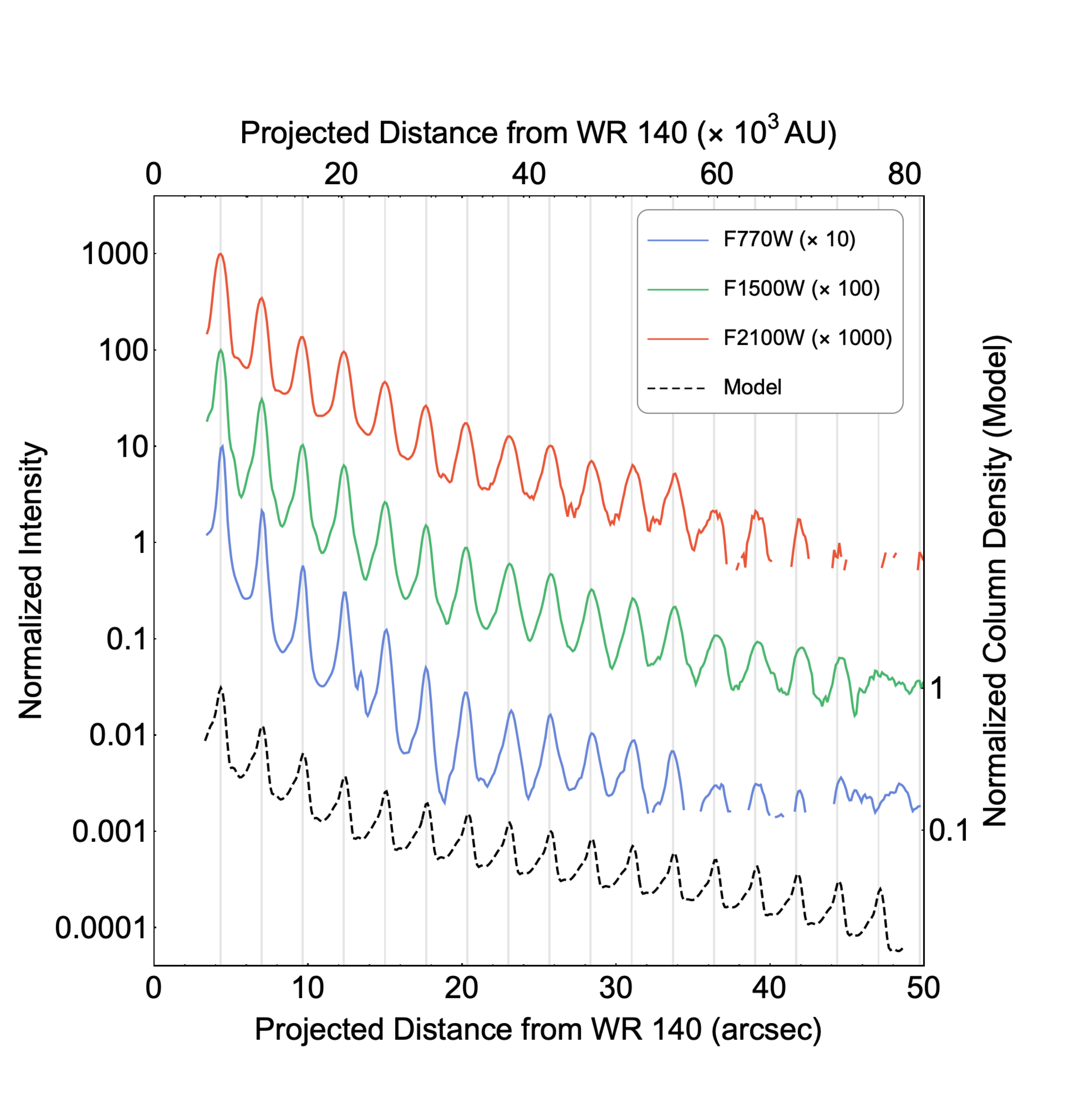}}
    \caption{\textbf{Radial plots across C1 dust features from Imager observations and geometric dust model.} Radial profiles through the C1 dust features showing the 7.7, 15, and 21 $\mu$m (blue, green, and red lines) emission and the column density from the dust model (black, dashed line) normalized to the peak value at shell 2 and offset vertically for clarity. \rev{The lower bound of the emission is clipped at a level of $2\sigma$. Note that the radial profiles of dust emission and column density do not exactly match because thermal dust emission is also regulated by dust temperature, which is not incorporated in the dust column density model}. Gray vertical lines indicate \rev{intervals evenly spaced by} $2.67''$ (\rev{4380} AU), which is the median separation between the flux density peaks.}
    \label{fig:WR140RadFlux}
\end{figure}

\rev{The 20-shell geometric model presents a striking} match \rev{to} the overall morphology of WR~140's nested dust shells \rev{revealed by the MIRI Imager} (Fig.~\ref{fig:WR140Mod}).  \rev{Notably,} the asymmetric and linear \rev{NW and SE features are present in the model at an angle and radial extent consistent with the linear features from the Imager observations. A comparison to a model MIRI F1500W PSF generated by WebbPSF \citep{Perrin2014} also demonstrates that the linear features do not align with the eight diffraction spikes (Fig.~\ref{fig:WR140Mod} \textit{Left}). Based on the orientation of the projected northwestern and southeastern edges of the single dust shell (Fig.~\ref{fig:WR140Orbit} \textit{Left}), the linear NW and SE linear dust features are likely associated with the projection of those edges from repeating dust shells.} 
\rev{The model also closely reproduces the repeating bright dust features as well as} the regular spacing \rev{between} the dust shells.
\rev{A radial plot of the dust column density from the model and the observed dust emission in the direction of C1 from shell 2 and beyond is shown in Figure~\ref{fig:WR140RadFlux} and demonstrates the consistency in the shell spacing.}

The regular spacing of the observed emission from 17 shells \rev{and the agreement with the geometric model} illustrates \rev{an apparently} freely expanding trajectory of the dust shells and WR~140's consistent episodic dust-production history over the past $\sim130$ years.
The median separation of the first 17 dust shells along C1
is $2.67\pm0.07''$, which corresponds to \rev{$4380\pm120$} AU assuming a distance of 1.64 kpc to WR~140 \citep{Rate2020,Thomas2021}.
The separation distance matches the expected separation given the 326 mas yr$^{-1}$ (2540 km s$^{-1}$) proper motion of C1 \citep{Williams2009} over WR~140's 7.93-yr orbital period. 
The \rev{projected} expansion velocity of the dust shells calculated from the median shell separations and the orbital period is $\sim2600$ km s$^{-1}$, which is consistent with the previously measured C1 proper motion. \rev{Since the geometric models indicate that the expansion direction of the C1 feature in 3D space is nearly perpendicular to our line of sight, the projected velocity of C1 should be consistent with the true dust expansion velocity \citep{Han2022}. The dust expansion velocity of $\sim2600$ km s$^{-1}$ is therefore} slightly less than the terminal velocity of the WC star \rev{wind} ($v_\infty=2860$ km s$^{-1}$, \citep{Williams1989}).

\begin{table}[t!]
\begin{center}
\begin{minipage}{0.55\textwidth}
\caption{\rev{Projected Distances of C1 Feature from WR~140}}\label{tab:rad}
\begin{tabular*}{\textwidth}{@{\extracolsep{\fill}}ccccc@{\extracolsep{\fill}}}
\toprule
Shell & $r_{C1}$& $\Delta r_{C1}$& $r_{C1}$ & $\Delta r_{C1}$  \\
 & (arcsec) & (arcsec) & (AU) & (AU)   \\
\midrule
 1 & 1.63  & - & 2670& -  \\
 2 & 4.32  &2.69 & 7090& 4410\\
 3 & 6.98  &2.66& 11440& 4360 \\
 4 & 9.62 &2.64 & 15780& 4330\\
 5 & 12.33 &2.71& 20220& 4440\\
 6 & 15.01 &2.68& 24610& 4390 \\
 7 & 17.66 &2.65& 28960& 4350  \\
 8 & 20.24 &2.58 & 33200& 4240  \\
 9 & 23.08 &2.84& 37850& 4650 \\
 10 & 25.76 &2.68& 42250& 4400 \\
 11 & 28.41 &2.65 & 46590& 4340 \\
 12 & 31.14 &2.73& 51070& 4480 \\
 13 & 33.78 &2.64& 55400& 4330  \\
 14 & 36.56 &2.78& 59960& 4560 \\
 15 & 39.18 &2.62& 64250& 4300\\
 16 & 41.98 &2.80& 68850& 4590  \\
 17 & 44.60 & 2.61& 73140& 4290\\
\botrule
\end{tabular*}
\footnotetext{C1 feature properties of the detected shells. $r_{C1}$ is the projected separation distance between C1 features and WR~140, which was calculated from the F1500W image for shells 2 -- 17 and the MRS observation for shell 1. $\Delta r_{C1}$ is the separation distance \rev{with respect to the adjacent inner dust shell} \revv{adopting a distance of 1.64 kpc \citep{Rate2020, Thomas2021}}.}
\end{minipage}
\end{center}
\end{table}

\rev{The apparently constant expansion velocity of the dust shells is notable given that dust around WR~140 should experience acceleration and deceleration by radiation pressure and drag forces, respectively \citep{Williams2009,Han2022}. Although beyond the scope of this work, the dust dynamics and the balance between radiation pressure and drag forces in the surrounding environment of WR~140 warrants further investigation.}
\rev{Given that the distant shells approach the signal-to-noise detection limits and retain a relatively constant separation distance (Fig.~\ref{fig:WR140RadFlux}, Tab.~\ref{tab:rad}), fainter and cooler dust shells likely exist beyond those detected by the MIRI Imager observations.}
\rev{The detection of distant shells from dust production $\sim130$ yr ago and the expected presence of more distant shells indicates that the dust formed by WR~140 can survive its harsh circumstellar environment and likely enriches the surrounding ISM. Follow-up observations with JWST of the local ISM and the surrounding $\sim10$ pc bubble that may have been carved out by WR~140 \citep{Arnal2001} could provide valuable insight on the ISM enrichment. }

\begin{figure}[t!]
\includegraphics[width=0.99\linewidth]{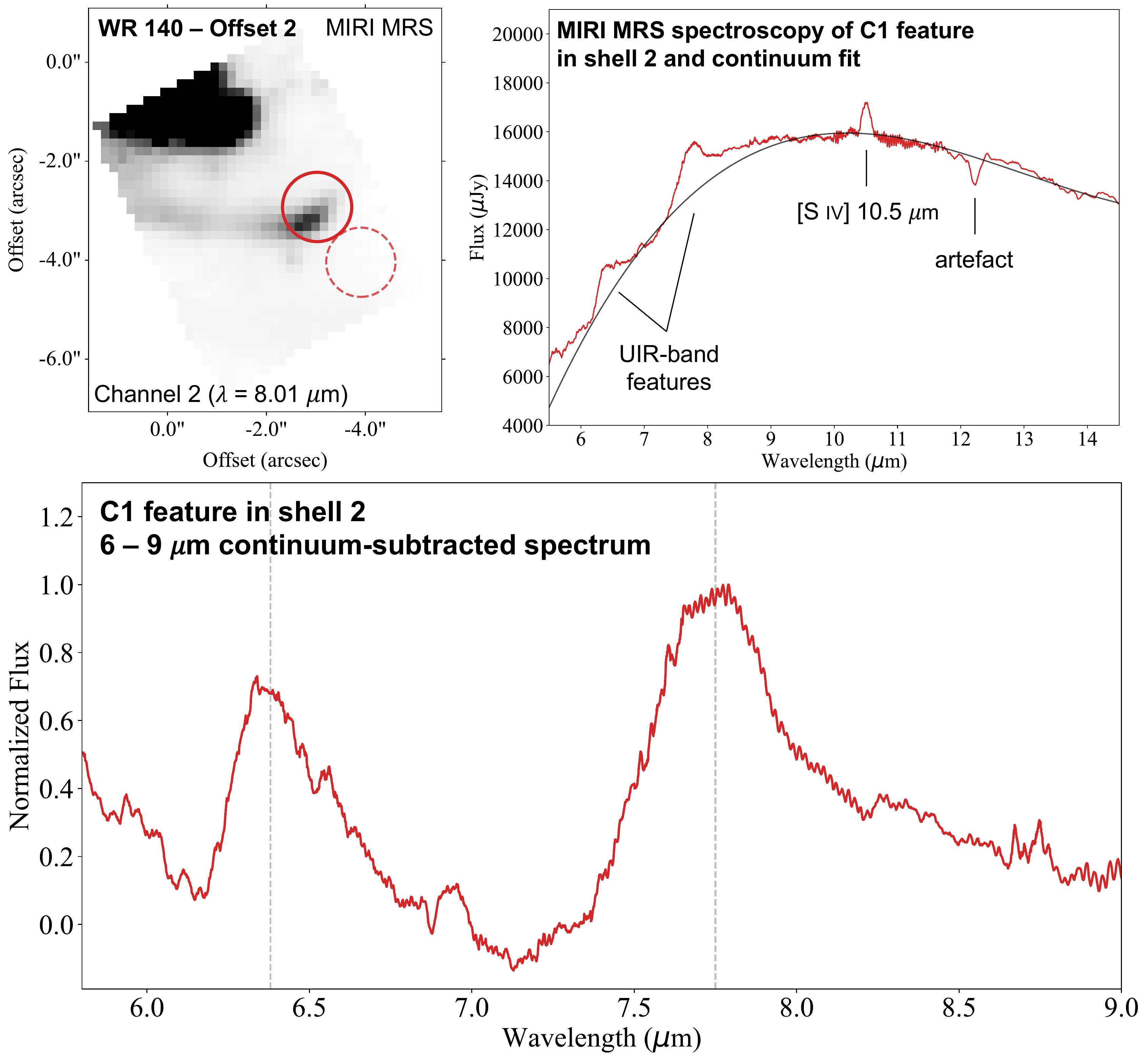}
    \caption{\textbf{MIRI MRS spectroscopy of the C1 feature in shell 2.} (\rev{\textit{Top Left}}) MRS spectral data cube image at 8.01 $\mu$m of WR~140 taken at the ``Offset 2'' position overlaid with the apertures used to extract the spectra from the C1 feature in shell 2 (solid) and measure the background emission (dashed). (\rev{\textit{Top Right}}) \rev{Extinction-corrected and background-subtracted} 6 -- 14 $\mu$m spectra of the C1 feature in shell 2 (red line) overlaid with the continuum fit from a \rev{cubic Chebyshev polynomial function} (black line). The absorption ``feature'' near 12~$\mu$m is an instrumental artefact, and the \rev{emission} feature at 10.5 $\mu$m is likely emission from the [S {\sc iv}] 10.5 $\mu$m fine-structure line. (\rev{\textit{Bottom}}) 6 -- 9 $\mu$m continuum-subtracted and normalized spectrum of the C1 feature from shell 2 exhibiting prominent 6.4 and 7.7 $\mu$m UIR-band features. Gaussian-fitted peaks of these features are located at $6.38$ $\mu$m and $7.75$ $\mu$m and are indicated by dashed vertical lines. Spectra were smoothed by convolving with a box filter kernel of width 21 spectral elements. }
    \label{fig:WR140Spec}
\end{figure}

\rev{Understanding the chemical properties and spectral signatures of dust formed by WC binaries like WR~140 is important given their potential role as dust sources in the ISM (e.g. \citep{Marchenko2017,Lau2020a,Lau2020b,Lau2021}). }
We can \rev{now} investigate \rev{this} using spatially and spectrally \rev{resolved observations of WR~140's circumstellar dust with the MIRI MRS}. The 6 -- 9~$\mu$m  wavelength interval includes a wealth of features that trace species of interstellar molecules and dust (see \citep{Peeters2002}), like Polycyclic Aromatic Hydrocarbons (PAHs), which are believed to be carriers of the ``unidentified infrared" (UIR) bands most commonly observed at 3.3, 6.2, 7.7, 11.2, and 12.7 $\mu$m \citep{LP84, allamandola85, Tielens08}. 
PAHs are known to be highly stable due to the honeycomb-like structure of carbon atoms fused in aromatic rings \citep{Allamandola1989, Tielens08}. The survival of WR~140's circumstellar \rev{dust} population in the harsh radiation field around the central binary would therefore be consistent with a composition of C-rich aromatic compounds.
UIR bands peaking at 6.4 and 7.9~-~8.5~$\mu$m have previously been reported from other dust-forming WC stars \citep{Chiar2002,Marchenko2017,Endo2022}; however, previous spectroscopic observations were unable to spatially resolve circumstellar dust emission that blended with a bright central core. \revvv{Mid-IR IFU spectroscopy of WR~140 with MIRI MRS can therefore} be used to confirm the association of the UIR features with the WC dust.

Figure~\ref{fig:WR140Spec} presents the spectral data cube image of WR~140 at 8.01~$\mu$m\rev{, the extinction-corrected spectrum of C1 from shell 2 along with the dust-continuum fit (See Sec.~\ref{sec:contsub}),} and the 6 -- 9~$\mu$m dust-continuum-subtracted and normalized spectrum from the shell 2 C1 feature. 
We identify the presence of two prominent UIR bands at 6.4 and 7.7~$\mu$m, which resemble the features from previous unresolved spectroscopic observations of other dust-forming WC stars. 
The peak positions of these features are at $6.38$ $\mu$m and $7.75$ $\mu$m based on a simple Gaussian fit to the continuum-subtracted spectrum \rev{(Fig.~\ref{fig:WR140Spec} \textit{Bottom})}.
The 7.7~$\mu$m feature is typically attributed to aromatic C-C stretching modes in PAHs \citep{Tielens08}. The 6.4~$\mu$m feature, which is primarily detected in H-poor environments from episodic dust-forming sources such as novae \citep{Helton2011}, R~Coronae~Borealis stars \citep{GarciaHernandez2013}, and other WC stars \citep{Endo2022}, is thought to originate from large carbonaceous molecules or small, H-free carbonaceous grains \citep{Harrington1998,Chiar2002}. The detection of the 6.4~$\mu$m band from circumstellar dust around WR~140 is therefore consistent with its expected H-poor environment dominated by the C-rich WC-star wind. The common 8.6 $\mu$m UIR feature, which is attributed to C--H bending modes, is notably absent from the spectrum and is also consistent \rev{with} an H-poor environment.
The direct detection of UIR bands from shell 2 indicates that \rev{dust formed by} WR~140 is carbonaceous and aromatic-rich. The stability of such aromatic compounds \citep{Allamandola1989,Tielens08} supports the interpretation of \rev{WC dust} survival out to the distant shells as revealed by the Imager observations (Fig.~\ref{fig:WR140Image}~\&~\ref{fig:WR140RadFlux}).

Due to their prevalence throughout the Universe and their influence on the thermal regulation and chemistry of the ISM (see \citep{Li2020} and ref.~therein), carriers of these UIR-band features are important components of the ISM.
The dominant source of UIR-band carriers has widely been thought to be evolved, solar-mass asymptotic giant branch (AGB) stars \citep{Allamandola1989,Galliano2008}.
However, the timescale for massive stars to evolve to the WR phase ($\sim$Myr) is much shorter than the timescale for lower mass stars to evolve into AGB stars ($\gtrsim100$ Myr).
\rev{Based on a dust-production analysis with Binary Population and Spectral Synthesis (BPASS, \citep{Eldridge2017}) models by \citep{Lau2020a}, the population of WC binaries can notably produce as much dust as the population of AGB stars depending on star-formation histories and metallicities.}
The detection of the C1 feature out to distant shells (e.g. shell \rev{17}; Fig.~\ref{fig:WR140RadFlux}) also \rev{demonstrates} that \rev{the carbonaceous WC dust} persists in the luminous and hard radiation field of the central binary system for at least \rev{130} yr after the initial dust-formation event \rev{and can likely propagate to the surrounding ISM.}
We therefore argue that dust-forming WC binaries like WR~140 could be considered \rev{as a} possible early and potentially dominant source of organic compounds and carbonaceous \rev{dust} in the ISM of our Galaxy and galaxies beyond.

\section{Methods}

\subsection{JWST Mid-Infrared Instrument Observations}
\label{sec:Obs}

\begin{table}[h]
\caption{\textbf{WR~140 MIRI Observation Summary.}}
\begin{center}
\begin{tabular}{lll}
\hline
\multicolumn{1}{l}{} & \multicolumn{1}{c}{MIRI MRS} & \multicolumn{1}{c}{MIRI Imager} \\
\hline 
Coordinates (J2000) & 20:20:27.985, +43:51:15.90 & 20:20:27.976, +43:51:16.28 \\
   &   20:20:27.781, +43:51:13.05   & \\
Observation Date (UTC) & 2022 Jul 8 (MJD 59768) & 2022 Jul 27 (MJD 59787)   \\
WR~140 Phase ($\varphi$) & 0.70 & 0.71 \\
Channels/Filters & Channels 1 -- 4 & F770W, F1500W, F2100W \\
Wavelength ($\mu$m)& \multicolumn{1}{l}{Ch 1: 4.9--7.65}  &  7.7, 15.0, 21.0 \\
& \multicolumn{1}{l}{Ch 2: 7.51--11.71}  &\\
& \multicolumn{1}{l}{Ch 3: 11.55--18.03} & \\
& \multicolumn{1}{l}{Ch 4: 17.71--28.1}   &  \\
Spec.~Resolving Power~($\lambda/\Delta \lambda$) & 1500 -- 3500 & 0.25, 0.48, 0.67 \\
Spaxel/Pixel Scale ($''$) & 0.13--0.35 & 0.11 \\
Field of View & $3.2''\times 2.7''$  -- $6.6''\times 7.6''$  & $112.6''\times 73.5''$ \\
\hline
\end{tabular}
\end{center}
\footnotetext{\revv{Observation summary table of the JWST/MIRI MRS and Imager observations. The two sets of MRS coordinates correspond to the pointings of the Offset 1 and Offset 2 observations (Fig.~\ref{fig:WR140Image}, \textit{Right}), and the Imager coordinate indicates the pointing of the Imager observation. The parameter $\varphi$ indicates WR~140's orbital phase at the time corresponding to the MRS and Imager observations based on the orbit described by \citep{Thomas2021}. The IFU channels and wavelength ranges used in the MRS observations are provided as well as the Imager filters and their corresponding central wavelengths. The spectral resolving power, spaxel/pixel scale, and field of view of both observing modes are also indicated.}}
\label{tab:Obs}
\end{table}%

\subsubsection{MIRI Imager}
\label{sec:Imager}
WR~140 (J2000 R.A. 20:20:27.98, Dec. +43:51:16.29; \citep{Lindegren2021}) was observed with the JWST Mid-Infrared Instrument (MIRI) Imager on 2022 July 27 UT with the F770W, F1500W, and F2100W filters in the FULL subarray mode. The Imager observations in each filter were performed using 3 sets of 4-Point dithers at starting set 5 in the positive direction with the default pattern size that was optimized for an extended source. The FASTR1 readout pattern with 43 groups, 1 integration, and 1 exposure per dither was used for each filter observation. The total exposure time in each filter was 0.40 hr. A summary of the MIRI Imager observations of WR~140 is provided in Tab.~\ref{tab:Obs}.

The MIRI Imager data were processed with the JWST calibration pipeline version 1.6.2 and the STCal version 1.0.0. After running the detector level pipeline the rate file was recreated from the rateints file to ensure the correct rate value was reported in regions that saturate in after the third group. We note that at the time of reducing this dataset, a dedicated non-linearity correction for the F2100W data was not available. Since the MIRI detectors exhibit a wavelength-dependency in their non-linear behavior, dedicated corrections are needed at wavelengths redder than 18 $\mu$m. We therefore expect an additional error contribution of few percent at 21 $\mu$m.

\subsubsection{MIRI Medium Resolution Spectroscopy}
\label{sec:MRS}
WR~140 was observed with the MIRI Medium-Resolution Spectrometer (MRS, \citep{Wells2015}) on 2022 July 8 UT. MIRI MRS observations provided spatially resolved spectroscopy over the four channels (Ch1 -- 4) that cover the 4.9 -- 28.1 $\mu$m with a spectral resolving power of $R\sim1500-3500$ \citep{Labiano2021}. Two sets of MRS observations were performed that targeted the two most recent dust shells formed by WR~140 5.5 and 13.4 yr ago (Fig.~\ref{fig:WR140Image}, \textit{Right}). The coordinates of the MRS observations and the other observing details are provided in Tab.~\ref{tab:Obs}. The two MRS observations were linked in a non-interruptible sequence due to the high proper motion of the circumstellar dust shells ($\sim300$ mas yr$^{-1}$, \citep{Williams2009}).

The first MRS observation utilized the bright core of WR~140 for target acquisition, which used the neutral density filter (FND) for 11.1 sec in the FAST readout pattern. Target acquisition was not used for the 2nd observation in the non-interruptible sequence, given the small offset ($\sim$4 arcsec) required from the initial position.
Each MRS observation was performed using the 4-Point dither optimized for an extended source and oriented in the negative direction. 60 groups and 5 integrations were used for each dither position and corresponded to 0.94 hr of exposure time in each grating position (SHORT, MEDIUM, LONG).  Simultaneous imaging with the MIRI Imager was also utilized in the MRS observations to improve the astrometric accuracy of the MRS data.

Calibration level 1 MIRI MRS data of WR~140 were obtained from the Mikulski Archive for Space Telescopes (MAST). These level 1 data were then processed through the JWST calibration pipeline version 1.6.0, where ``output\_type = band'' was set in the ``cube\_build'' step to output 12 data cubes corresponding to the three grating positions of each four channel.

\subsection{Data Analysis}

\subsubsection{\rev{MIRI Imager Background Subtraction, }Radial Dust Emission Profile, and C1 Positions}

The median background emission in relatively empty patches of sky beyond the circumstellar shells in the F770W, F1500W, and F2100W MIRI Imager observation exhibited fluxes of 10.4, 41.4, and 210.2 MJy sr$^{-1}$, respectively. A background subtraction using the median values was therefore applied to the three MIRI Imager observations. 

A radial 3-pixel-width line cut plot originating at the position of WR~140 and oriented $53^\circ$ south of west was used to measure the emission and positions of the C1 features from the MIRI Imager observations (Fig.~\ref{fig:WR140RadFlux}; Tab.~\ref{tab:rad}). A Gaussian profile was used to measure the peak positions of each of the shells 2 -- 17 from the radial dust emission profile of the F1500W image. An identical line cut was used to measure the dust column density from the \rev{20-shell} geometric dust model.

\subsubsection{MIRI MRS Spectroscopy of C1}

Spectra were extracted from the MIRI MRS `Offset 1' observations of C1 in shell 1 (20:20:27.838, +43:51:14.69) using a circular aperture with a radius of $r=0.7''$. Subtraction of the background emission and contamination from the diffraction pattern of the bright core was performed by using a circular aperture in a relatively dust-free position NW of WR~140 (20:20:27.932, +43:51:17.75) at the same separation distance of the C1 aperture.  
However, contamination from emission lines from the bright stellar core appeared to persist in the shell 1 C1 spectra even after the background subtraction. 
Follow-up work from the WR DustERS program (ERS 1349) will be conducted to implement a technique to perform PSF subtraction on the MIRI MRS observations for a future study.

A $r=0.7''$ circular aperture was also used to extract the spectra from the MIRI MRS `Offset 2' observations of C1 in shell 2 (20:20:27.657, +43:51:12.970). Background emission for the C1 region in shell 2 was measured and subtracted using a circular background aperture with an identical radius centered on a relatively empty area in the field of view to the south-west of shell 2 at the following coordinates: 20:20:27.575, +43:51:11.852. The 6.4 and 7.7 $\mu$m features were present in background-subtracted spectra even when using different background positions in other relatively empty regions of the sky.

In order to mitigate high frequency noise from artefacts such as ``fringing,'' the MRS spectra were smoothed by convolving with a box filter kernel that had a width of 21 spectral elements ($\Delta\lambda=0.04$ $\mu$m). The width of the UIR features ($\gtrsim0.2$ $\mu$m) is notably larger than the width of the kernel used for smoothing.

\subsubsection{Spectral Stitching between MRS Channels}
In order to combine the spectra extracted from the 3 sub-bands from the four channels (Ch 1-4), the median flux density in the overlapping wavelengths between the 12 sub-bands was used to combine the extracted C1 spectra. The spectra were normalized to the Channel 2 ``SHORT'' sub-band ($\lambda=7.51-8.76$ $\mu$m), which overlaps with the $6-9$ $\mu$m wavelength range studied in this work.

\subsubsection{Dust-Continuum Subtraction}
\label{sec:contsub}
The dust continuum emission was \rev{fit and} subtracted from the $\sim6-9$ $\mu$m wavelength range of the extracted spectra of the C1 region from shell 2 using \rev{a cubic Chebyshev polynomial function from the} ``Spectrum1D" and ``fit\_continuum" functions \rev{in} the Python package \textit{specutils}~\citep{Earl2022}. Wavelengths between 6.8 -- 7.2 $\mu$m and 9--15 $\mu$m were used to fit the C1 continuum in shell 2. Ionized gas likely attributed to the [S {\sc iv}] 10.5 $\mu$m fine-structure line is also detected in the C1 spectra \rev{and will be investigated in future analysis}. The apparent absorption ``feature'' near 12 $\mu$m is an instrumental artefact.
The shorter wavelengths ($\lambda\lesssim 6$ $\mu$m) were omitted for the shell 2 dust continuum fit to avoid enhanced emission likely arising from a hotter dust component (\revv{Lau et al., submitted}). The C1 spectra from shell 2 and the dust continuum \rev{fit} are shown in \rev{Fig.~\ref{fig:WR140Spec} (\textit{Top Right})}. The $\sim6-14$ $\mu$m continuum-subtracted spectra were then normalized to the peak flux density of the 7.7 $\mu$m feature (Fig.~\ref{fig:WR140Spec}, \rev{\textit{Bottom}}). 

\subsubsection{Extinction Correction}
Extracted MIRI MRS \rev{spectra} from the circumstellar dust around WR~140 were dereddened using the extinction correction factors derived from the Gordon et al.~(2021) \citep{Gordon2021} average Galactic interstellar extinction curve ($R_V=3.17$) with $A_{K_s}=0.24$ \citep{Rate2020}.

\subsection{WR140 Geometric Models}
\label{sec:model}

The circumstellar dust structures around WR~140 shown in Fig.~\ref{fig:WR140Mod} (Left) were simulated based on the geometric model presented in \citep{Han2022}. The model assumes dust to be produced on a thin conical surface originating from the wind-shock interface and has been applied in previous studies of dust-forming colliding-wind binaries \citep{Callingham2019,Lau2020b,Han2020}. 
The geometry of the dust structure depends in part on the orbit of the binary, for which we adopt the well-constrained orbital parameters from \citep{Thomas2021}. The half opening angle of the conical dust surface is assumed to be 40$^\circ$ \citep{Marchenko2003,Williams2009,Han2022}, and the dust expansion velocity is assumed to be 2600 km s$^{-1}$, which is consistent with the velocity measured from the separation distance between the dust shells (Tab.~\ref{tab:rad}).  
The only parameter of the geometric model that deviates from the original model fitted to ground-based observations of a single shell from \citep{Han2022} is the dust expansion speed, being slightly larger than the expansion speed of 2450 km s$^{-1}$ fitted to data at an earlier orbital phase (0.592). This apparent discrepancy may be reflective of acceleration during the expansion of the dust plume found by \citep{Han2022}. 
The model was then convolved with the F1500W MIRI Imager PSF from the WebbPSF Python package \citep{Perrin2014}.
\rev{Note that because only the dust morphology is generated by the model the central binary itself does not appear.}

Observations of WR~140 indicate that dust does not form evenly over the orbit \citep{Williams2009,Han2022}. By fitting to the structures in one shell of dust observed by ground-based AO-assisted IR imaging,~\citep{Han2022} found that the structures are reproduced if dust production from the colliding-winds begins at a true anomaly of -135$^\circ$ and ceases at +135$^\circ$ (Fig.~\ref{fig:WR140Orbit}), between which dust production is suppressed. No dust is assumed to be produced along other parts of the orbit. Furthermore, the structures were better reproduced when assuming that the dust density peaks on the trailing edge of the conical shock front, becoming progressively lower when moving towards the leading edge. Further details of the geometric model of WR~140 are provided in the study by \citep{Han2022}.

\section{\revv{Code Availability Statement}}
\revv{This research made use of Astropy, a community-developed core Python package for Astronomy \citep{Astropy2013, Astropy2018}. This research also made use of Jdaviz, which is a package of astronomical data analysis visualization tools based on the Jupyter platform \citep{Lim2022}.}

\section{\revv{Data Availability}}
\revv{Data used in this study were obtained under the JWST DD-ERS program ERS 1349 and have no exclusive access period. Data can be obtained from the Mikulski Archive for Space Telescopes (MAST). }

\section{\revv{Competing Interests}}
\revv{The authors declare no competing interests.}

\section{\revv{Author Contributions}}
\revv{R.M.L. led the analysis and is PI of the WR DustERS Team. R.M.L. and M.J.H. conceived and designed the project. I.A. and D.L. processed the MRS data, and M.G.M. processed the MIRI imaging data. Y.H. and P.T. constructed the geometric models of WR~140. All authors contributed to observation planning and/or scientific interpretation as members of the WR DustERS Team.}

\section{Acknowledgements}
RML would like to acknowledge the members of our entire WR DustERS team for their valuable discussions and contributions to this work. 
We thank Amaya Moro-Martin, William Januszewki, Neill Reid, Margaret Meixner, and Bonnie Meinke for their support of the planning and execution of our ERS program. We would also like to acknowledge the MIRI instrument and \textit{MIRISim} teams for their insightful feedback and support of our observation and data analysis plans.
We are also grateful to Karl Gordon for his guidance on the MIRI Imager data reduction. 
\rev{Lastly, we would like to thank the anonymous reviewers for their prompt and valuable feedback that helped to improve the quality and focus of this work.}

The work of RML is supported by NOIRLab, which is managed by the Association of Universities for Research in Astronomy (AURA) under a cooperative agreement with the National Science Foundation.
YH acknowledges funding from the Gates Cambridge Trust.
MFC and KH were supported by NASA under award number 80GSFC21M0002.
OCJ acknowledges support from an STFC Webb fellowship.
AFJM is grateful for financial aid from NSERC (Canada).
J.S.-B. acknowledges the support received from the Mexican Council of Science (CONACyT) “Ciencia de Frontera” project CF-2019/263975.
\revv{CMPR acknowledges support from NATA ATP grant 80NSSC22K0628 and NASA Chandra Theory grant TM2-23003X.}

This work is based on observations made with the NASA/ESA/CSA James Webb Space Telescope. The data were obtained from the Mikulski Archive for Space Telescopes at the Space Telescope Science Institute, which is operated by the Association of Universities for Research in Astronomy, Inc., under NASA contract NAS 5-03127 for JWST. These observations are associated with program ERS1349.

Support for program ERS1349 was provided by NASA through a grant from the Space Telescope Science Institute, which is operated by the Association of Universities for Research in Astronomy, Inc., under NASA contract NAS 5-03127.


\begin{thebibliography}{10}
\expandafter\ifx\csname url\endcsname\relax
  \def\url#1{\burl{#1}}\fi
\expandafter\ifx\csname urlprefix\endcsname\relax\def\urlprefix{URL }\fi
\providecommand{\bibinfo}[2]{#2}
\providecommand{\eprint}[2][]{\url{#2}}
\providecommand{\doi}[1]{\url{https://doi.org/#1}}
\bibcommenthead

\bibitem{Marchenko2017}
\bibinfo{author}{{Marchenko}, S.~V.} \& \bibinfo{author}{{Moffat}, A.~F.~J.}
\newblock \bibinfo{title}{{Search for polycyclic aromatic hydrocarbons in the
  outflows from dust-producing Wolf-Rayet stars}}.
\newblock \emph{\bibinfo{journal}{\mnras}} \textbf{\bibinfo{volume}{468}},
  \bibinfo{pages}{2416--2428} (\bibinfo{year}{2017}).

\bibitem{Lau2020a}
\bibinfo{author}{{Lau}, R.~M.} \emph{et~al.}
\newblock \bibinfo{title}{{Revisiting the Impact of Dust Production from
  Carbon-rich Wolf-Rayet Binaries}}.
\newblock \emph{\bibinfo{journal}{\apj}} \textbf{\bibinfo{volume}{898}},
  \bibinfo{pages}{74} (\bibinfo{year}{2020}).

\bibitem{Endo2022}
\bibinfo{author}{{Endo}, I.} \emph{et~al.}
\newblock \bibinfo{title}{{Detection of a Broad 8 {\ensuremath{\mu}}m UIR
  Feature in the Mid-infrared Spectrum of WR 125 Observed with Subaru/COMICS}}.
\newblock \emph{\bibinfo{journal}{\apj}} \textbf{\bibinfo{volume}{930}},
  \bibinfo{pages}{116} (\bibinfo{year}{2022}).

\bibitem{Crowther2007}
\bibinfo{author}{{Crowther}, P.~A.}
\newblock \bibinfo{title}{{Physical Properties of Wolf-Rayet Stars}}.
\newblock \emph{\bibinfo{journal}{\araa}} \textbf{\bibinfo{volume}{45}},
  \bibinfo{pages}{177--219} (\bibinfo{year}{2007}).

\bibitem{Usov1991}
\bibinfo{author}{{Usov}, V.~V.}
\newblock \bibinfo{title}{{Stellar wind collision and dust formation in
  long-period, heavily interacting Wolf-Rayet binaries.}}
\newblock \emph{\bibinfo{journal}{\mnras}} \textbf{\bibinfo{volume}{252}},
  \bibinfo{pages}{49} (\bibinfo{year}{1991}).

\bibitem{Williams2009}
\bibinfo{author}{{Williams}, P.~M.} \emph{et~al.}
\newblock \bibinfo{title}{{Orbitally modulated dust formation by the WC7+O5
  colliding-wind binary WR140}}.
\newblock \emph{\bibinfo{journal}{\mnras}} \textbf{\bibinfo{volume}{395}},
  \bibinfo{pages}{1749--1767} (\bibinfo{year}{2009}).

\bibitem{Eatson2022}
\bibinfo{author}{{Eatson}, J.~W.}, \bibinfo{author}{{Pittard}, J.~M.} \&
  \bibinfo{author}{{Van Loo}, S.}
\newblock \bibinfo{title}{{Exploring dust growth in the episodic WCd system
  WR140}}.
\newblock \emph{\bibinfo{journal}{arXiv e-prints}}
  \bibinfo{pages}{arXiv:2204.12354} (\bibinfo{year}{2022}).

\bibitem{Tuthill1999}
\bibinfo{author}{{Tuthill}, P.~G.}, \bibinfo{author}{{Monnier}, J.~D.} \&
  \bibinfo{author}{{Danchi}, W.~C.}
\newblock \bibinfo{title}{{A dusty pinwheel nebula around the massive star
  WR104}}.
\newblock \emph{\bibinfo{journal}{\nat}} \textbf{\bibinfo{volume}{398}},
  \bibinfo{pages}{487--489} (\bibinfo{year}{1999}).

\bibitem{Cherchneff2000}
\bibinfo{author}{{Cherchneff}, I.}, \bibinfo{author}{{Le Teuff}, Y.~H.},
  \bibinfo{author}{{Williams}, P.~M.} \& \bibinfo{author}{{Tielens},
  A.~G.~G.~M.}
\newblock \bibinfo{title}{{Dust formation in carbon-rich Wolf-Rayet stars. I.
  Chemistry of small carbon clusters and silicon species}}.
\newblock \emph{\bibinfo{journal}{\aap}} \textbf{\bibinfo{volume}{357}},
  \bibinfo{pages}{572--580} (\bibinfo{year}{2000}).

\bibitem{Monnier2011}
\bibinfo{author}{{Monnier}, J.~D.} \emph{et~al.}
\newblock \bibinfo{title}{{First Visual Orbit for the Prototypical
  Colliding-wind Binary WR 140}}.
\newblock \emph{\bibinfo{journal}{\apjl}} \textbf{\bibinfo{volume}{742}},
  \bibinfo{pages}{L1} (\bibinfo{year}{2011}).

\bibitem{Fahed2011}
\bibinfo{author}{{Fahed}, R.} \emph{et~al.}
\newblock \bibinfo{title}{{Spectroscopy of the archetype colliding-wind binary
  WR 140 during the 2009 January periastron passage}}.
\newblock \emph{\bibinfo{journal}{\mnras}} \textbf{\bibinfo{volume}{418}},
  \bibinfo{pages}{2--13} (\bibinfo{year}{2011}).

\bibitem{Thomas2021}
\bibinfo{author}{{Thomas}, J.~D.} \emph{et~al.}
\newblock \bibinfo{title}{{The orbit and stellar masses of the archetype
  colliding-wind binary WR 140}}.
\newblock \emph{\bibinfo{journal}{\mnras}} \textbf{\bibinfo{volume}{504}},
  \bibinfo{pages}{5221--5230} (\bibinfo{year}{2021}).

\bibitem{Rate2020}
\bibinfo{author}{{Rate}, G.} \& \bibinfo{author}{{Crowther}, P.~A.}
\newblock \bibinfo{title}{{Unlocking Galactic Wolf-Rayet stars with Gaia DR2 -
  I. Distances and absolute magnitudes}}.
\newblock \emph{\bibinfo{journal}{\mnras}} \textbf{\bibinfo{volume}{493}},
  \bibinfo{pages}{1512--1529} (\bibinfo{year}{2020}).

\bibitem{Bouchet2015}
\bibinfo{author}{{Bouchet}, P.} \emph{et~al.}
\newblock \bibinfo{title}{{The Mid-Infrared Instrument for the James Webb Space
  Telescope, III: MIRIM, The MIRI Imager}}.
\newblock \emph{\bibinfo{journal}{\pasp}} \textbf{\bibinfo{volume}{127}},
  \bibinfo{pages}{612} (\bibinfo{year}{2015}).

\bibitem{Wright2015}
\bibinfo{author}{{Wright}, G.~S.} \emph{et~al.}
\newblock \bibinfo{title}{{The Mid-Infrared Instrument for the James Webb Space
  Telescope, II: Design and Build}}.
\newblock \emph{\bibinfo{journal}{\pasp}} \textbf{\bibinfo{volume}{127}},
  \bibinfo{pages}{595} (\bibinfo{year}{2015}).

\bibitem{Wells2015}
\bibinfo{author}{{Wells}, M.} \emph{et~al.}
\newblock \bibinfo{title}{{The Mid-Infrared Instrument for the James Webb Space
  Telescope, VI: The Medium Resolution Spectrometer}}.
\newblock \emph{\bibinfo{journal}{\pasp}} \textbf{\bibinfo{volume}{127}},
  \bibinfo{pages}{646} (\bibinfo{year}{2015}).

\bibitem{Lau2020b}
\bibinfo{author}{{Lau}, R.~M.} \emph{et~al.}
\newblock \bibinfo{title}{{Resolving Decades of Periodic Spirals from the
  Wolf-Rayet Dust Factory WR 112}}.
\newblock \emph{\bibinfo{journal}{\apj}} \textbf{\bibinfo{volume}{900}},
  \bibinfo{pages}{190} (\bibinfo{year}{2020}).

\bibitem{Han2020}
\bibinfo{author}{{Han}, Y.} \emph{et~al.}
\newblock \bibinfo{title}{{The extreme colliding-wind system Apep: resolved
  imagery of the central binary and dust plume in the infrared}}.
\newblock \emph{\bibinfo{journal}{\mnras}} \textbf{\bibinfo{volume}{498}},
  \bibinfo{pages}{5604--5619} (\bibinfo{year}{2020}).

\bibitem{Han2022}
\bibinfo{author}{{Han}, Y.}, \bibinfo{author}{{Tuthill}, P.~G.},
  \bibinfo{author}{{Lau}, R.~M.} \& \bibinfo{author}{{Soulain}, A.}
\newblock \bibinfo{title}{{Radiation driven acceleration in the expanding WR140
  dust shell}}.
\newblock \bibinfo{note}{{Accepted to Nature}}.

\bibitem{Perrin2014}
\bibinfo{author}{{Perrin}, M.~D.} \emph{et~al.}
\newblock \bibinfo{editor}{{Oschmann}, J., Jacobus~M.},
  \bibinfo{editor}{{Clampin}, M.}, \bibinfo{editor}{{Fazio}, G.~G.} \&
  \bibinfo{editor}{{MacEwen}, H.~A.} (eds) \emph{\bibinfo{title}{{Updated point
  spread function simulations for JWST with WebbPSF}}}.
\newblock (eds \bibinfo{editor}{{Oschmann}, J., Jacobus~M.},
  \bibinfo{editor}{{Clampin}, M.}, \bibinfo{editor}{{Fazio}, G.~G.} \&
  \bibinfo{editor}{{MacEwen}, H.~A.}) \emph{\bibinfo{booktitle}{Space
  Telescopes and Instrumentation 2014: Optical, Infrared, and Millimeter
  Wave}}, Vol. \bibinfo{volume}{9143} of \emph{\bibinfo{series}{Society of
  Photo-Optical Instrumentation Engineers (SPIE) Conference Series}},
  \bibinfo{pages}{91433X} (\bibinfo{year}{2014}).

\bibitem{Williams1989}
\bibinfo{author}{{Williams}, P.~M.} \& \bibinfo{author}{{Eenens}, P.~R.~J.}
\newblock \bibinfo{title}{{Displaced He I absorption lines in Wolf-Rayet stars
  : revisions to v-infinite.}}
\newblock \emph{\bibinfo{journal}{\mnras}} \textbf{\bibinfo{volume}{240}},
  \bibinfo{pages}{445--457} (\bibinfo{year}{1989}).

\bibitem{Arnal2001}
\bibinfo{author}{{Arnal}, E.~M.}
\newblock \bibinfo{title}{{A High-Resolution H I Study of the Interstellar
  Medium Local to HD 193793}}.
\newblock \emph{\bibinfo{journal}{\aj}} \textbf{\bibinfo{volume}{121}},
  \bibinfo{pages}{413--425} (\bibinfo{year}{2001}).

\bibitem{Lau2021}
\bibinfo{author}{{Lau}, R.~M.} \emph{et~al.}
\newblock \bibinfo{title}{{Revealing Efficient Dust Formation at Low
  Metallicity in Extragalactic Carbon-rich Wolf-Rayet Binaries}}.
\newblock \emph{\bibinfo{journal}{\apj}} \textbf{\bibinfo{volume}{909}},
  \bibinfo{pages}{113} (\bibinfo{year}{2021}).

\bibitem{Peeters2002}
\bibinfo{author}{{Peeters}, E.} \emph{et~al.}
\newblock \bibinfo{title}{{The rich 6 to 9 vec mu m spectrum of interstellar
  PAHs}}.
\newblock \emph{\bibinfo{journal}{\aap}} \textbf{\bibinfo{volume}{390}},
  \bibinfo{pages}{1089--1113} (\bibinfo{year}{2002}).

\bibitem{LP84}
\bibinfo{author}{{Leger}, A.} \& \bibinfo{author}{{Puget}, J.~L.}
\newblock \bibinfo{title}{{Identification of the Unidentified Infrared Emission
  Features of Interstellar Dust}}.
\newblock \emph{\bibinfo{journal}{\aap}} \textbf{\bibinfo{volume}{137}},
  \bibinfo{pages}{L5--L8} (\bibinfo{year}{1984}).

\bibitem{allamandola85}
\bibinfo{author}{{Allamandola}, L.~J.}, \bibinfo{author}{{Tielens},
  A.~G.~G.~M.} \& \bibinfo{author}{{Barker}, J.~R.}
\newblock \bibinfo{title}{{Polycyclic aromatic hydrocarbons and the
  unidentified infrared emission bands: auto exhaust along the milky way.}}
\newblock \emph{\bibinfo{journal}{\apjl}} \textbf{\bibinfo{volume}{290}},
  \bibinfo{pages}{L25--L28} (\bibinfo{year}{1985}).

\bibitem{Tielens08}
\bibinfo{author}{{Tielens}, A.~G.~G.~M.}
\newblock \bibinfo{title}{{Interstellar polycyclic aromatic hydrocarbon
  molecules.}}
\newblock \emph{\bibinfo{journal}{\araa}} \textbf{\bibinfo{volume}{46}},
  \bibinfo{pages}{289--337} (\bibinfo{year}{2008}).

\bibitem{Allamandola1989}
\bibinfo{author}{{Allamandola}, L.~J.}, \bibinfo{author}{{Tielens},
  A.~G.~G.~M.} \& \bibinfo{author}{{Barker}, J.~R.}
\newblock \bibinfo{title}{{Interstellar Polycyclic Aromatic Hydrocarbons: The
  Infrared Emission Bands, the Excitation/Emission Mechanism, and the
  Astrophysical Implications}}.
\newblock \emph{\bibinfo{journal}{\apjs}} \textbf{\bibinfo{volume}{71}},
  \bibinfo{pages}{733} (\bibinfo{year}{1989}).

\bibitem{Chiar2002}
\bibinfo{author}{{Chiar}, J.~E.}, \bibinfo{author}{{Peeters}, E.} \&
  \bibinfo{author}{{Tielens}, A.~G.~G.~M.}
\newblock \bibinfo{title}{{The Infrared Emission Features in the Spectrum of
  the Wolf-Rayet star WR 48a}}.
\newblock \emph{\bibinfo{journal}{\apjl}} \textbf{\bibinfo{volume}{579}},
  \bibinfo{pages}{L91--L94} (\bibinfo{year}{2002}).

\bibitem{Helton2011}
\bibinfo{author}{{Helton}, L.~A.}, \bibinfo{author}{{Evans}, A.},
  \bibinfo{author}{{Woodward}, C.~E.} \& \bibinfo{author}{{Gehrz}, R.~D.}
\newblock \bibinfo{editor}{{Joblin}, C.} \& \bibinfo{editor}{{Tielens},
  A.~G.~G.~M.} (eds) \emph{\bibinfo{title}{{Atypical dust species in the ejecta
  of classical novae}}}.
\newblock (eds \bibinfo{editor}{{Joblin}, C.} \& \bibinfo{editor}{{Tielens},
  A.~G.~G.~M.}) \emph{\bibinfo{booktitle}{EAS Publications Series}},
  Vol.~\bibinfo{volume}{46} of \emph{\bibinfo{series}{EAS Publications
  Series}}, \bibinfo{pages}{407--412} (\bibinfo{year}{2011}).

\bibitem{GarciaHernandez2013}
\bibinfo{author}{{Garc{\'\i}a-Hern{\'a}ndez}, D.~A.}, \bibinfo{author}{{Rao},
  N.~K.} \& \bibinfo{author}{{Lambert}, D.~L.}
\newblock \bibinfo{title}{{Dust Around R Coronae Borealis Stars. II. Infrared
  Emission Features in an H-poor Environment}}.
\newblock \emph{\bibinfo{journal}{\apj}} \textbf{\bibinfo{volume}{773}},
  \bibinfo{pages}{107} (\bibinfo{year}{2013}).

\bibitem{Harrington1998}
\bibinfo{author}{{Harrington}, J.~P.}, \bibinfo{author}{{Lame}, N.~J.},
  \bibinfo{author}{{Borkowski}, K.~J.}, \bibinfo{author}{{Bregman}, J.~D.} \&
  \bibinfo{author}{{Tsvetanov}, Z.~I.}
\newblock \bibinfo{title}{{Discovery of a 6.4 Micron Dust Feature in
  Hydrogen-Poor Planetary Nebulae}}.
\newblock \emph{\bibinfo{journal}{\apjl}} \textbf{\bibinfo{volume}{501}},
  \bibinfo{pages}{L123--L126} (\bibinfo{year}{1998}).

\bibitem{Li2020}
\bibinfo{author}{{Li}, A.}
\newblock \bibinfo{title}{{Spitzer's perspective of polycyclic aromatic
  hydrocarbons in galaxies}}.
\newblock \emph{\bibinfo{journal}{\nastro}} \textbf{\bibinfo{volume}{4}},
  \bibinfo{pages}{339--351} (\bibinfo{year}{2020}).

\bibitem{Galliano2008}
\bibinfo{author}{{Galliano}, F.}, \bibinfo{author}{{Dwek}, E.} \&
  \bibinfo{author}{{Chanial}, P.}
\newblock \bibinfo{title}{{Stellar Evolutionary Effects on the Abundances of
  Polycyclic Aromatic Hydrocarbons and Supernova-Condensed Dust in Galaxies}}.
\newblock \emph{\bibinfo{journal}{\apj}} \textbf{\bibinfo{volume}{672}},
  \bibinfo{pages}{214--243} (\bibinfo{year}{2008}).

\bibitem{Eldridge2017}
\bibinfo{author}{{Eldridge}, J.~J.} \emph{et~al.}
\newblock \bibinfo{title}{{Binary Population and Spectral Synthesis Version
  2.1: Construction, Observational Verification, and New Results}}.
\newblock \emph{\bibinfo{journal}{\pasa}} \textbf{\bibinfo{volume}{34}},
  \bibinfo{pages}{e058} (\bibinfo{year}{2017}).

\bibitem{Lindegren2021}
\bibinfo{author}{{Lindegren}, L.} \emph{et~al.}
\newblock \bibinfo{title}{{Gaia Early Data Release 3. The astrometric
  solution}}.
\newblock \emph{\bibinfo{journal}{\aap}} \textbf{\bibinfo{volume}{649}},
  \bibinfo{pages}{A2} (\bibinfo{year}{2021}).

\bibitem{Labiano2021}
\bibinfo{author}{{Labiano}, A.} \emph{et~al.}
\newblock \bibinfo{title}{{Wavelength calibration and resolving power of the
  JWST MIRI Medium Resolution Spectrometer}}.
\newblock \emph{\bibinfo{journal}{\aap}} \textbf{\bibinfo{volume}{656}},
  \bibinfo{pages}{A57} (\bibinfo{year}{2021}).

\bibitem{Earl2022}
\bibinfo{author}{{Earl}, N.} \emph{et~al.}
\newblock \bibinfo{title}{{astropy/specutils: V1.7.0}}.
\newblock \bibinfo{howpublished}{Zenodo} (\bibinfo{year}{2022}).

\bibitem{Gordon2021}
\bibinfo{author}{{Gordon}, K.~D.} \emph{et~al.}
\newblock \bibinfo{title}{{Milky Way Mid-Infrared Spitzer Spectroscopic
  Extinction Curves: Continuum and Silicate Features}}.
\newblock \emph{\bibinfo{journal}{\apj}} \textbf{\bibinfo{volume}{916}},
  \bibinfo{pages}{33} (\bibinfo{year}{2021}).

\bibitem{Callingham2019}
\bibinfo{author}{{Callingham}, J.~R.} \emph{et~al.}
\newblock \bibinfo{title}{{Anisotropic winds in a Wolf-Rayet binary identify a
  potential gamma-ray burst progenitor}}.
\newblock \emph{\bibinfo{journal}{\nastro}} \textbf{\bibinfo{volume}{3}},
  \bibinfo{pages}{82--87} (\bibinfo{year}{2019}).

\bibitem{Marchenko2003}
\bibinfo{author}{{Marchenko}, S.~V.} \emph{et~al.}
\newblock \bibinfo{title}{{The Unusual 2001 Periastron Passage in the
  ``Clockwork'' Colliding-Wind Binary WR 140}}.
\newblock \emph{\bibinfo{journal}{\apj}} \textbf{\bibinfo{volume}{596}},
  \bibinfo{pages}{1295--1304} (\bibinfo{year}{2003}).

\bibitem{Astropy2013}
\bibinfo{author}{{Astropy Collaboration}} \emph{et~al.}
\newblock \bibinfo{title}{{The Astropy Project: Building an Open-science
  Project and Status of the v2.0 Core Package}}.
\newblock \emph{\bibinfo{journal}{\aj}} \textbf{\bibinfo{volume}{156}},
  \bibinfo{pages}{123} (\bibinfo{year}{2018}).

\bibitem{Astropy2018}
\bibinfo{author}{{Astropy Collaboration}} \emph{et~al.}
\newblock \bibinfo{title}{{The Astropy Project: Building an Open-science
  Project and Status of the v2.0 Core Package}}.
\newblock \emph{\bibinfo{journal}{\aj}} \textbf{\bibinfo{volume}{156}},
  \bibinfo{pages}{123} (\bibinfo{year}{2018}).

\bibitem{Lim2022}
\bibinfo{author}{{Lim}, P.~L.} \emph{et~al.}
\newblock \bibinfo{title}{{spacetelescope/jdaviz: v2.8.0}}.
\newblock \bibinfo{howpublished}{Zenodo} (\bibinfo{year}{2022}).

\end{thebibliography}



\end{document}